\documentclass[runningheads]{llncs}
\usepackage[T1]{fontenc}
\usepackage[misc]{ifsym}
\usepackage{graphicx}
\usepackage{color}
\usepackage{adjustbox}
\usepackage{amssymb}

\usepackage{tabularx}
\newcommand{\headrow}[1]{\multicolumn{1}{c}{\adjustbox{angle=45,lap=\width-0.5em}{#1}}}

\usepackage{xspace}
\newcommand{\etal}{\textit{et al.}\xspace}
\newcommand{\etc}{\textit{etc.}\xspace}
\newcommand{\ie}{\textit{i.e.,}\xspace}
\newcommand{\eg}{\textit{e.g.,}\xspace}
\newcommand{\cf}{\textit{cf.}\xspace}

\begin{document}

\title{Quest Love: A First Look at Blockchain Loyalty Programs}
\titlerunning{Quest Love}

\author{
	Joseph Al-Chami\inst{1} \and 
	Jeremy Clark\inst{2}\orcidID{0000-0002-3533-5965}
	}

\institute{
	Independent Researcher \\ \email{alchamijoseph@gmail.com} \and
	Concordia University, Montreal, Canada \\ \email{j.clark@concordia.ca} 
	}

\maketitle


\begin{abstract}

Blockchain ecosystems---such as those built around chains, layers, and services---try to engage users for a variety of reasons: user education, growing and protecting their market share, climbing metric-measuring leaderboards with competing systems, demonstrating usage to investors, and identifying worthy recipients for newly created tokens (airdrops). A popular approach is offering user quests: small tasks that can be completed by a user, exposing them to a common task they might want to do in the future, and rewarding them for completion. In this paper, we analyze a proprietary dataset from one deployed quest system that offered 43 unique quests over 10 months with 80M completions. We offer insights about the factors that correlate with task completion: amount of reward, monetary value of reward, difficulty, and cost. We also discuss the role of farming and bots, and the factors that complicate distinguishing real users from automated scripts.

\end{abstract}




\section{Introductory Remarks}

Blockchain projects understand that even the best technical ideas and code may struggle to have an impact without user adoption of the platform. There is no clear method for reaching pseudonymous users, who rarely seek out new blockchain platforms, and there’s no prominent `DApp store.' 

One approach for introducing users to a new blockchain (including layer 1 chains, sidechains, and layer 2 solutions) is showcasing its features and partners through gamification. A set of \textit{quests} or \textit{tasks} or \textit{boosts} are created that offer a \textit{bounty} or \textit{reward} to users for performing a specific action on the blockchain, such as transferring a token or using a specific DeFi service. The reward could be a collectible badge (often as an NFT), points on a leaderboard (often referred to as XP, inspired by games like Fortnite and platforms like Duolingo), or an airdrop of tokens with some monetary value~\cite{FB19,MYL24,YL24}. Most projects favour the term `quests' over `tasks' to avoid associations with employment. The term quest highlights the user's journey toward achievement and mastery, creating a more legally favourable framing rooted in gamification. Despite the aspirational tone, these actions remain structured and transactional, serving both users’ reward-driven motivations and projects’ growth and engagement objectives.

A few examples of quest initiatives include Arbitrum's Odyssey 2.0, Avalanche's Coachella Quests, Base's Builder Quests, Binance's BSC GameFi Quest and Polygon's Wallet Suite Quest. Platforms like Galxe, Layer3, Boost, DeBank, and Zealy have emerged as centralized hubs offering users a variety of quests from multiple projects and have adopted the description: the loyalty layer. 

In this measurements paper, we study one quest system over time to understand how the completion-rate of quests is correlated with factors such as difficulty, cost, reward amount, and reward type (monetary or non-monetary value). From our data, we are not able to directly classify users between humans and bots, however we do offer some insights and discussion points about the influence of automated task-completion. To inform this discussion, we identify key stakeholders involved (such as blockchain platforms, blockchain-based services, investors, activity-monitoring services, users, and bots) and analyze their preferences and incentives. We maintain a neutral position on whether quests are useful or not---we study them because they are commonly used (perhaps poised for broader adoption) and we failed to find another detailed study on the parameters that correlate with task completion. 


\subsection{Research Questions} 
Our paper sheds light on the following research question (RQ) and two discussion questions (DQs). DQs have less (or no) empirical basis and are instead based on insights or discussion points.

\begin{itemize}
\item \textbf{RQ1:} What factors are correlated with quest completion rates? We analyze (a) reward amount, (b) task difficulty, (c) cost, (d) reserve requirements, and (e) rewards with monetary value.
\item \textbf{DQ1:} To what extent can the design of the quest help distinguish genuine users from automated bots and farming?
\item \textbf{DQ2:} Who are the stakeholders in quest systems? What can we learn from their preferences?
\end{itemize}


\subsection{Related Work}

We are not aware of other research papers on quest systems for blockchain projects. Brand loyalty programs~\cite{DU97} have been studied deeply, and more recently gamification of learning tasks~\cite{HKS14}. Experience points in games have been criticized~\cite{Bog13}. For loyalty programs, Wulf \etal outline several drivers of participation~\cite{Wulf03}. Relevant to our study are: users engage when costs are zero as opposed to non-zero, they are asked to make minimal efforts as opposed to extended efforts, and when rewards are `hard' (\eg a discount) as opposed to `soft' (\eg product updates). 

On the blockchain side, the relationship between new tokens and user adoption has been studied~\cite{BH19}. One particular (and recent) focus is on airdrops~\cite{FB19,MYL24,YL24} where a large quantity (majority to totality) of tokens are distributed to users for the first time. This literature finds a positive impact of airdrops on token value, that retaining user engagement after an airdrop is difficult, and evidence of artificial engagement to `farm' airdropped tokens (including detection algorithms~\cite{FTWC23,ZCH+24}). In our dataset, the quest system is run both before and after an airdrop. In discussing the role of farming and bots, we use a persuasion model~\cite{Fogg09} which suggests certain deterrents to engagement: time, money, physical effort, brain cycles, social deviance, and non-routine. We examine the relationship between quest difficulty (time, brain cycles, and non-routine) and cost (money) with completion, assuming most quests have similar levels of physical effort (minimal) and social deviance (legal or unregulated in most jurisdictions). 


\subsection{Dataset \& Limitations}

\paragraph{Reproducibility.} For our paper, we study a quest system built and maintained by a company offering a new EVM-compatible, proof of stake blockchain. The quests themselves were co-designed with various services (DApps) running on the company's blockchain. The choice to use proprietary data has pros and cons. 

The disadvantage is that we offered the company final say over being named or identifiable in our paper; they declined after our study was complete. Thus we only provide our dataset after processing and de-identifying.\footnote{Available on GitHub: \url{https://github.com/MadibaGroup/2025-Quests}} Our statistics and figures can be replicable from the data and new insights could be developed by other researchers, however the accuracy of the data cannot be independently verified against on-chain data or website archives.

The advantage of engaging with a company is that we were able to inform the quest design to some extent. For example, we were able to suggest the addition of reserve requirements (§\ref{sec:thresh}) to quests and rewards with monetary value (§\ref{sec:rewards}), and ensure they were added/removed from tasks that otherwise were unchanged (isolating two main variables: the requested feature and the epoch). To control for different completion rates across epochs, we also ensured some tasks were offered across multiple epochs without change.

\paragraph{Representativeness.} Our dataset originates from a single platform, which means we cannot assert broad generalizability. However, the platform's quests and bot-control measures appear standard and comparable to other blockchain quest systems we examined. Our data covers both pre- and post-airdrop periods, enabling us to observe user behaviour changes around significant events. It's important to note that our findings may become less representative over time due to the increasing ease with which non-experts can create bots, particularly as AI-driven agents~\cite{Ante24,WGN+25} become more prevalent.

\paragraph{Ground Truth on Bots.} We lack ground truth distinguishing genuine users from Sybil accounts, whether manual (farming) or automated (bots/scripts). Thus, our data reports only unique address completions, and interpretations should be made cautiously of this fact. Nevertheless, understanding factors correlated with quest completion rates remains valuable for designing effective blockchain engagement strategies, even without explicit bot classification.


\subsection{Data Analysis}

The data we used includes the quest system's website interface where each quest corresponds to specific actions executed on-chain, defined by interactions with smart contracts. Upon quest completion, the loyalty program verifies the user's on-chain transaction to confirm that all quest requirements are met. Once verified, the program stores the details in the company's database—including transaction hash, user address, and contract address. This information is merged with additional off-chain data such as the points awarded for quest completion and the epoch, a time metric controlled by the loyalty program rather than the blockchain. We were provided SQL access to this raw data and exported CSV files (anonymized and available on our repository) for analysis. The statistics we compute (such as Spearman correlation) are documented in our repository. We skip the details in the paper as the techniques themselves not novel or innovative. 


\subsection{Key Events in Dataset}

Our research analyzes data spanning the entirety of 2024. New quests are offered for an epoch, which was initially a one day period and transitioned (at epoch 10) to one week. Our data is organized by epoch 5--42 (the company's backend database starts at epoch 5). Over this period, 43 unique quests were offered. Some quests were offered in multiple epochs and could be completed multiple times. 1.2M unique addresses completed at least one quest.  Across the period, 80M quests were completed.

\begin{figure}[t]
    \centering
    \includegraphics[width=0.8\textwidth]{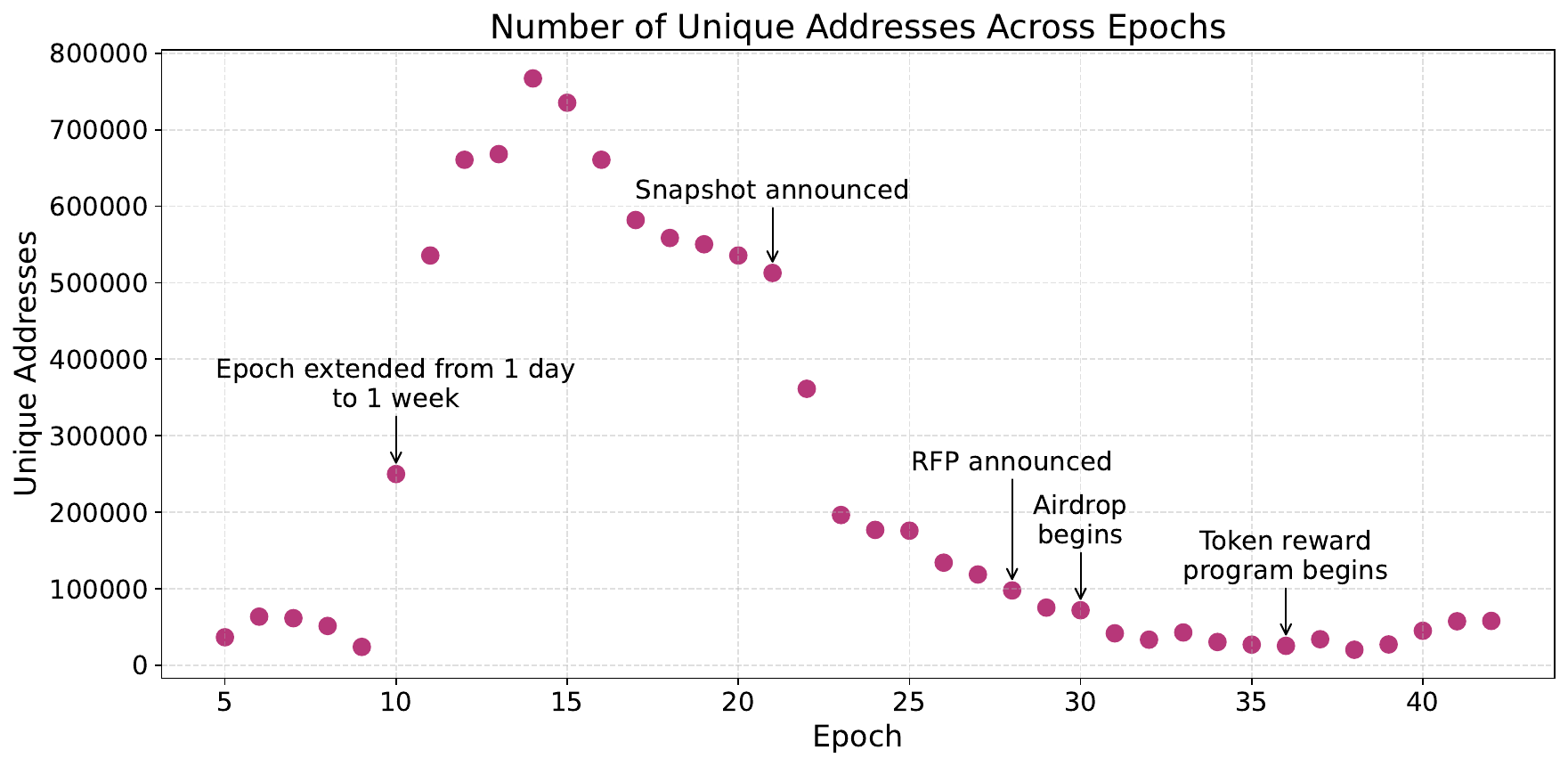}
    \caption{Total completions of all quests offered in each epoch. Overlay of public announcements that may impact user participation.}
    \label{fig:events}
\end{figure}

As seen in  Figure~\ref{fig:events}, overall participation seems to be influenced by various public announcements. When a blockchain does not offer a token but does offer quests, speculative users might assume that an airdrop of tokens is on the blockchain's future roadmap and quest completion points might factor into who receives airdropped tokens. At epoch 21, the project announces an airdrop and that a snapshot of user activity had been captured as a basis for the airdrop. Since the snapshot is taken, it is too late to complete quests for the purposes of gaining in airdropped tokens (unless you speculate that the project will do a second snapshot or a second airdrop). Participation dropped 29.5\% immediately after the snapshot announcement. By epoch 23, a further 45.7\% drop occurred, totalling a 61.7\% decline over two weeks. At epoch 28, a request for proposals (RFP) was announced to ask ecosystem apps and partners to submit how they want to do airdrop allocation, which further cements the fact that quest completions after the snapshot will not be considered. A 22.9\% drop in participation follows the RFP. At epoch 30, the RFP closed and the airdrop itself starts, which sees another 42\% decline in quest completions. In summary, from epoch 21 to epoch 32, quest completions declined 93.5\%.  

In epoch 36, a new program was launched where certain quests would receive token rewards (which have monetary value), as opposed to just receiving XP points. We will study this further later but for now, we note that this leads an increase in participation: 34\% increase in epoch 37 and 128.7 \% by epoch 42. The reward program kept evolving and increasing rewards over the following epochs. By epoch 50 the cumulative effect of this token program was 251.8\%.

Retention rate can be computed as $RR= \frac{(E-N)}{S}\cdot 100$ for users at the start ($S$), end ($E$), and newly acquired users ($N$). Between epochs 21 and 35, RR was -0.183\%. After token rewards were introduced (epoch 42), $RR$ rose slightly to just 0.59\%. Transaction volume dropped sharply from 11,301,078 at peak participation to just 572,595 by epoch 39 (5\% of peak volume). Taken together, retention is extremely low and incentive-driven (reacting to monetary rewards). This supports the idea that user engagement was transactional and profit-driven rather than loyalty-driven.


\section{RQ1: Task Completion Factors}
\subsection{Reward Amounts in Points}

\begin{figure}[t]
    \centering
    \includegraphics[width=0.8\textwidth]{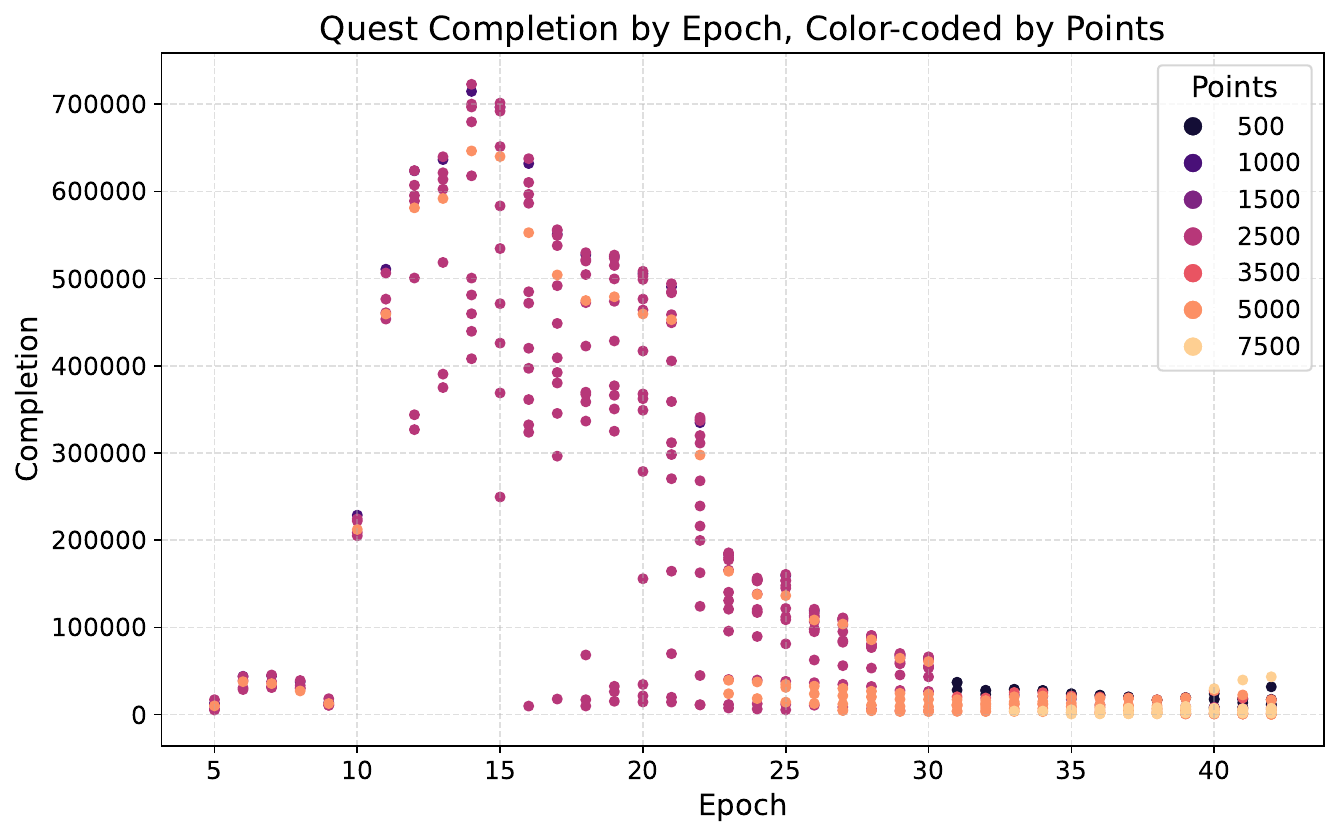}
    \caption{Reward amount (in non-monetary points) of different quests offered from Epoch 5--42. Completion rates do not correlate with points, indicating other contributing factors.\label{fig:points}}
\end{figure}

Points have long been a key element in gamification systems~\cite{HKS14}, primarily used to incentivize specific behaviours or actions, such as educational tasks, objectives in a game, or fostering loyalty through purchases at a business. For quests, points are usually presented in leaderboards---a public, ordered ranking of users and their points, which can boost competition, and result in users striving not merely to collect points but to climb the ranks. Leaderboards have a downside too. Newcomers might feel deterred by the seemingly insurmountable gap between themselves and top-ranked users, and excessive focus on ranking can foster cheating or the transacting of user profiles (with their points).

To what extent do points motivate users? Consider Figure~\ref{fig:points} where it shows a dot for each quest offered in each epoch (x-axis) and the number of times it was completed (y-axis). The shape of the curve follows the overall trend (Figure~\ref{fig:events}). The dots in Figure~\ref{fig:points} are colour-coded for how many points the user receives for completing the quest. A simple hypothesis is that quest completions would correlate positively with the number of points. Using the Spearman correlation~\cite{Spe04}, we found that correlations across epochs ranged from -0.115 to -0.870, with no epoch showing a significant positive monotonic relationship. In other words, the rank ordering of points and completions does not reveal any consistent patterns—for instance, higher points do not reliably correspond to more completions.

Finally, we note that in the later epochs, a strong negative monotonic relationship emerged, indicating that higher-point quests were completed less frequently. This was an intentional design decision, as more challenging quests were assigned higher point rewards to incentivize completion. As the data suggests, higher point rewards alone were not associated with increased engagement for these harder quests. Instead, quest difficulty, and other factors that act as deterrents, are more strongly associated with variations in task completion. This finding highlights the need to consider additional factors to better understand the predictors of task completion.


\subsection{Task Difficulty}

\begin{figure}[t]
    \centering
    \includegraphics[width=0.8\textwidth]{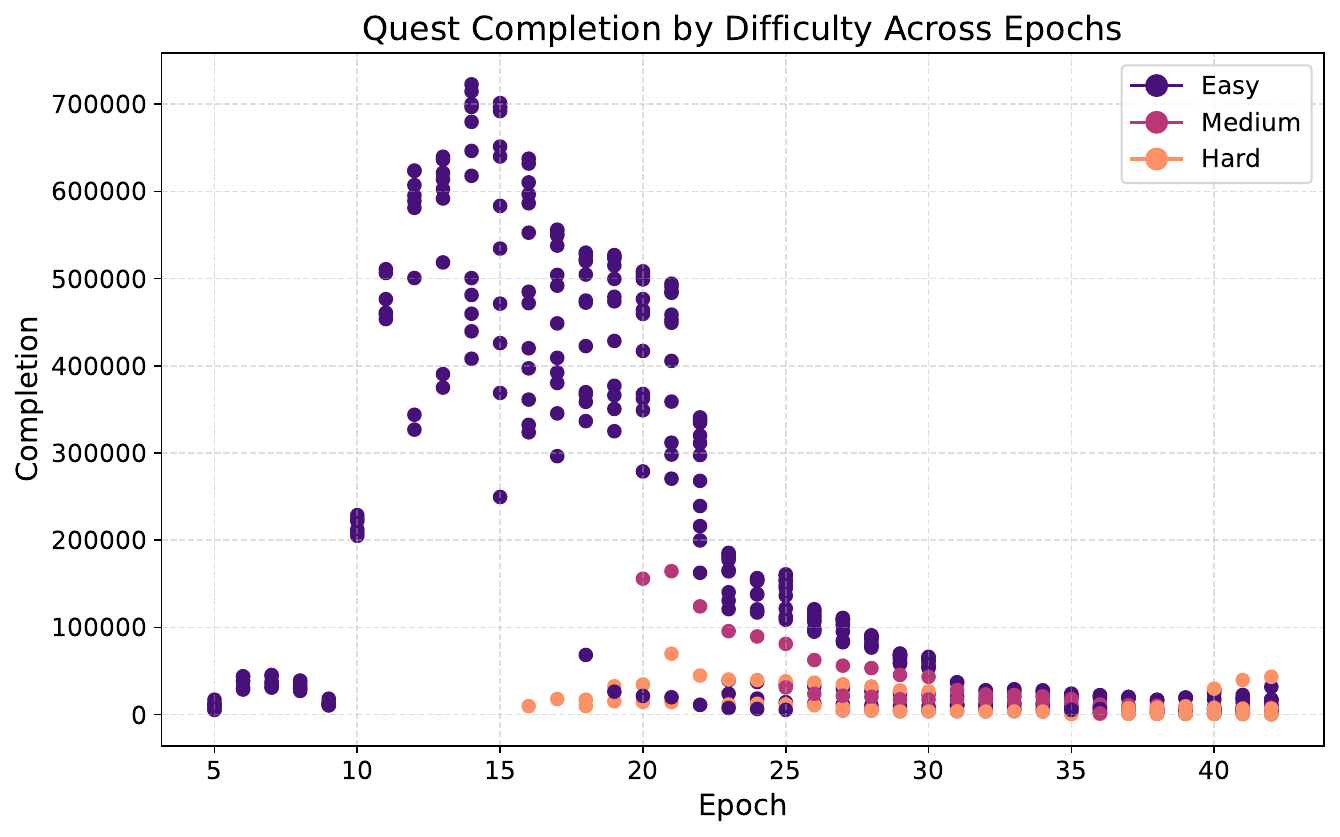}
    \caption{Quest difficulty is strongly associated with completion rates, with easy quests (purple) completed most often and difficult quests (orange) completed least often. Variance in the right tail (epochs 40+) correspond to offering medium quests (pink) with tokens of monetary value.\label{fig:difficulty}}
\end{figure}

The first research question addresses points, which pull users toward certain quests. On the other side, frictions or deterrents may push users away from certain quests. We willl address three of these: difficulty, cost, and minimum token requirements (or thresholds), starting with task difficulty. We classify all quests based on (i) completion time and (ii) external dependencies that are out of the user's direct control, such as waiting for a response from another system, interacting with other users, or relying on network availability. Completion time was measured by lead researcher who completed each quest modelling average user behaviour; we felt this was objective enough but more robust followup studies might take the agreement of several researchers independently assessing difficulty.

\begin{itemize}
\item Easy: quest takes less than 1 minute and has no external dependencies.
\item Medium: quest takes 1--2 minutes and/or has external dependencies but dependencies require minimal effort to resolve.
\item Hard: quest takes longer than 2 minutes and/or has significant external dependencies.
\end{itemize}

An example of an easily resolved external dependency would be finding another player for a game where many players are available at all times. By contrast, if players were rarely available, the quest would be ranked hard instead of medium. If a quest falls between two levels, we will typically classify it as the easier of the two. 

As evident from Figure~\ref{fig:difficulty}, difficulty is strongly associated with task completion. Using Spearman correlation (with reversed difficulty mapping), we found that correlations across epochs ranged from 0.121 to 0.663, mostly indicating a positive monotonic relationship. Easier tasks consistently had higher completion rates. 

Nevertheless, in certain epochs, the positive correlation was not as strong. As shown in Figure~\ref{fig:difficulty}, some medium or hard quests were completed more frequently than easier tasks, suggesting that other deterrents or negative incentives may be more strongly associated with quest completion than difficulty alone. This observation is further explored by analyzing factors such as cost and thresholds in subsequent sections.


\subsection{Cost}

\begin{figure}[t]
    \centering
    \includegraphics[width=0.8\textwidth]{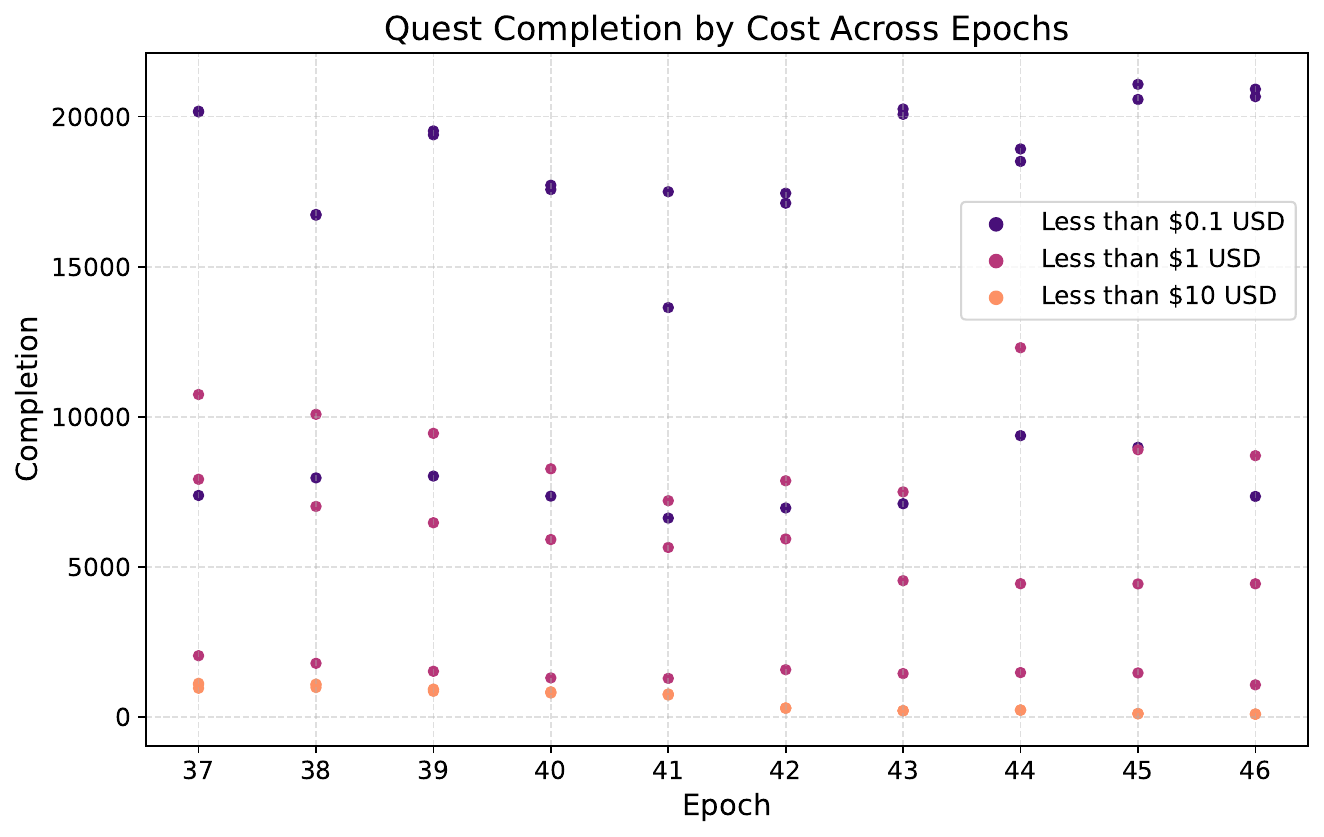}
    \caption{Quest cost is strongly associated with completion rates, with cheaper quests (purple) completed most often and more expensive quests (orange) completed least often.\label{fig:cost}}
\end{figure}

All quests involving an on-chain transaction will have non-zero cost to complete, as the user needs to pay the transaction fee. The more complex the task, the larger the gas fee could be. For example, if a quest involves using a DeFi service, the service itself may charge fees on top of the gas costs. Quests that involve bridging assets between (layer 1) Ethereum and the blockchain project seemed particularly disliked, likely due to gas fees being high on Ethereum. 

In Figure~\ref{fig:cost}, we show a selection of quests from epochs 37--46 (later quests demonstrated more variety in complexity) and coded their cost into three colours: purple is less than \$0.10 USD, pink is less than \$1.00 USD (and more than purple), and orange is \$1.00 USD and more (no quest was greater than \$10 USD but some cost greater than \$5.00 USD). Quest costs in USD were computed using token exchange rates at the time of each transaction, based on price data provided by CoinGecko, aligning precisely with the prices displayed to users on the quest platform UI. The analysis of quest completion data revealed a clear trend: quests with a cost under \$0.10 USD are consistently completed by a larger number of users across all epochs, while those costing between \$1 and \$10 USD see minimal engagement. This pattern suggests that cost is strongly associated with lower engagement in high-cost quests. Given this observed relationship, quest designers may benefit from considering alternative incentives or cost structures.

Using the Spearman correlation, we found that correlations across epochs ranged from -0.843 to -0.896, consistently showing the strongest monotonic relationship among all research questions.
As shown in Figure 3, the rank ordering of cost and completions reveals a clear pattern—less costly quests (e.g., purple, costing less than \$0.10) are completed far more frequently than higher-cost quests (e.g., orange, costing up to \$10). This finding reinforces the idea that cost is strongly associated with lower task completion.

Since cost showed the strongest correlation, we infer that users are not completing quests randomly, nor are they primarily motivated by mastery or loyalty. Instead, their behavior suggests a tendency to minimize cost, aligning with patterns commonly associated with Sybils or automated bots. This observation led us to the next research question, where thresholds were introduced to user wallets to investigate their association with completion rates and assess whether financial deterrents influence engagement in a more structured manner.


\subsection{Thresholds}
\label{sec:thresh}

\begin{figure}[t]
    \centering
    \includegraphics[width=0.8\textwidth]{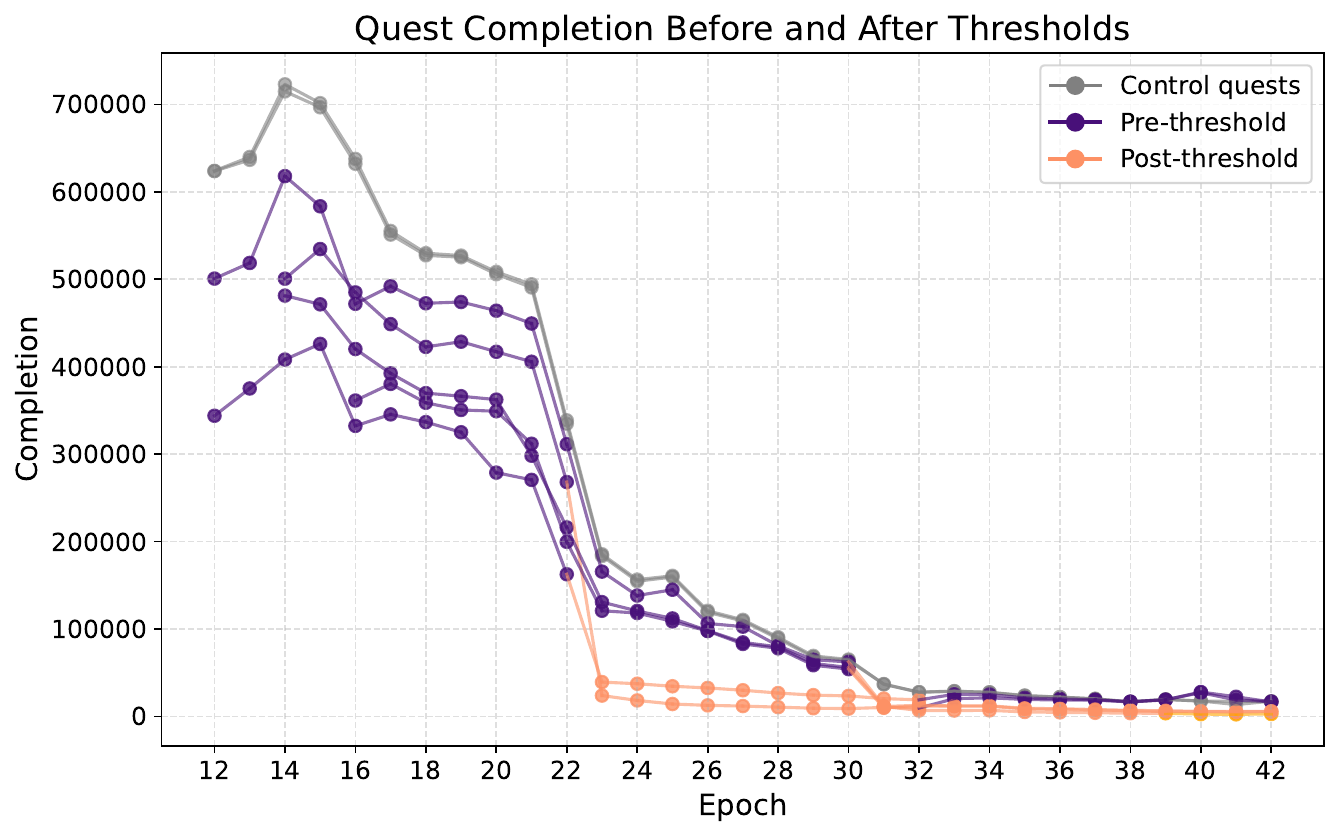}
    \caption{The implementation of a minimum threshold of tokens for quest completion is associated with an immediate decline in completions for quests that previously ran without thresholds (orange segment compared to the purple segment). The grey lines represent highly completed tasks as a control group.\label{fig:threshold}}
\end{figure}

We consider one last complication for users: quests that can only be completed by accounts with a minimum threshold of tokens. Thresholds are used to combat farming and bots, which we discussed further in the next section. For now, we simply measure their correlation with quest completion. 

Consider Figure~\ref{fig:threshold}. It shows 5 quests (purple lines) that ran for multiple epochs without a threshold, and then a threshold was applied to them (purple lines turn orange). In a few cases, the threshold was removed and the line goes back to purple. As a control group of quests for comparison, the grey line shows 5 highly completed quests (all simple token transfers). Throughout all epochs, these tasks had no threshold requirements for completion, and no thresholds were added at any point during the measurement. Most thresholds were in the native token on the blockchain project, and the amount of tokens required were worth between \$1 and \$10 USD. Some thresholds were in small amounts of ETH (0.002) or BTC (0.000125). 

As apparent from the figure, the implementation of a threshold is associated with an immediate decline in quest completion. The purple lines follow the same trend as the grey (control) lines (dampened by not being easy tasks) until the threshold is implemented, and then plummet. One might hypothesize that bots and farmers would take a few epochs to update their scripts or acquire threshold tokens, but the association appears sticky with no future uptick. When the threshold is removed, the quests' completion rates return to levels similar to those observed in the control group.

For example, one quest was completed 268,104 times in epoch 22, just before a threshold of approximately \$10 USD was introduced at epoch 23. Following the enforcement of the threshold, completions dropped dramatically to 23,923— a 91\% decrease. When the threshold was removed at epoch 33, the quest saw a significant rebound, with completions more than doubling, increasing by 112.8\%.

Similarly, another quest, which had a threshold of approximately \$5 USD enforced at epoch 23, experienced a 75.8\% drop in completions. After the threshold was removed at epoch 33, completions increased by 33.4\%. 

These examples highlight a substantial association between thresholds and lower user participation, as well as an association between the removal of thresholds and a return to higher completion rates.


\subsection{Rewards with Monetary Value}
\label{sec:rewards}

\begin{figure}[t]
    \centering
    \includegraphics[width=0.8\textwidth]{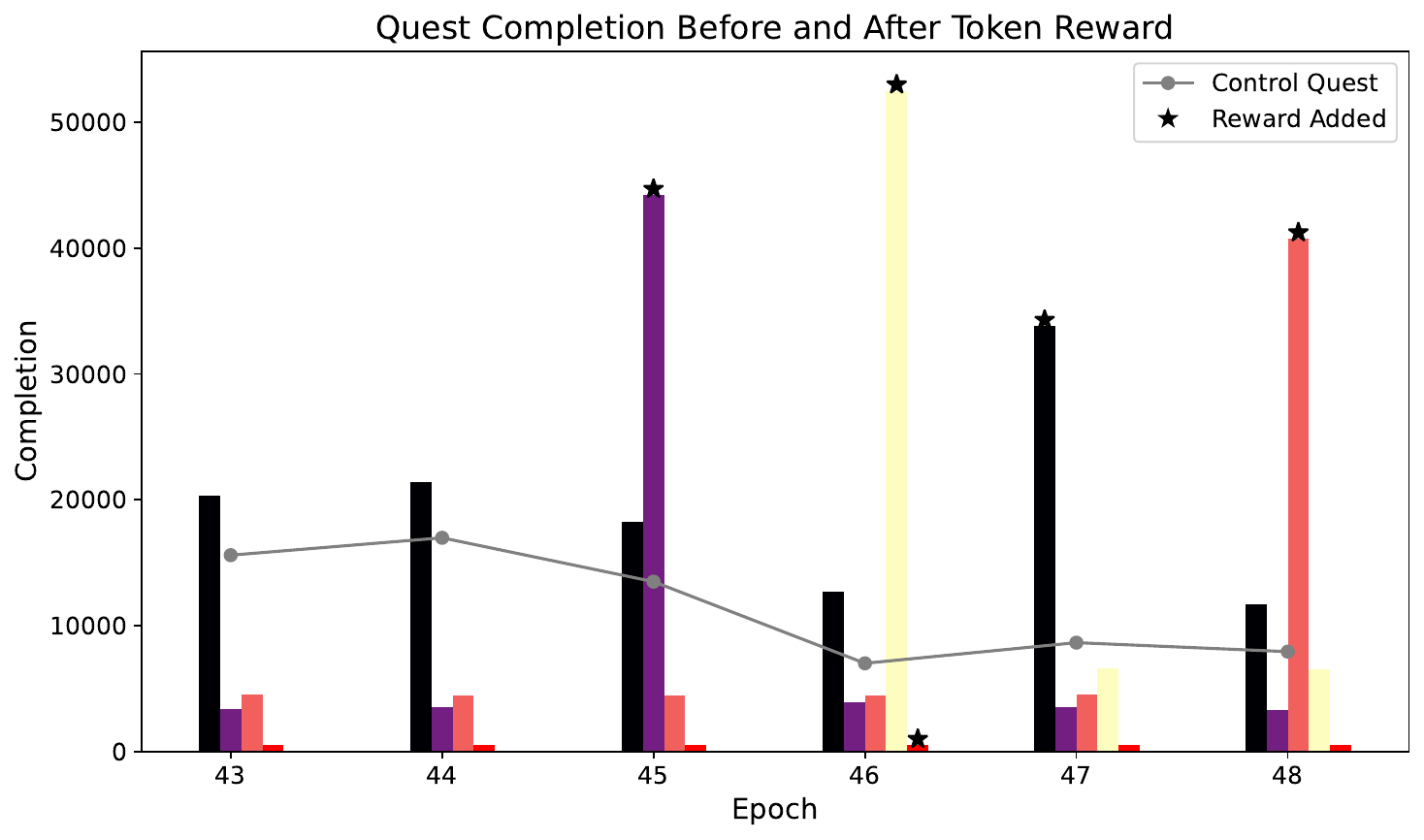}
    \caption{Five quests (by colour) with rewards in non-monetary points (bars without stars) and token rewards for one epoch (bars with stars). The grey line shows an average of task completion for the other quests over the same epochs. The chart shows that offering tokens (with a monetary value) as rewards is associated with higher quest completion. \label{fig:tokens}}
\end{figure}

Consider Figure~\ref{fig:tokens} which depicts 5 quests by colour, each offered over epochs 43--49 (except the yellow quest, which was first offered in epoch 46). For most epochs, the completion award was non-monetary (loyalty or experience) points, as we have analyzed in the previous section. In contrast, for one epoch (shown with a star), alongside the usual point rewards, specific quests were selected for a token reward (in the native token of the blockchain project) which had a small monetary value of ~\$0.10 USD. The grey line shows the completion rate of the control quest offered across the same epochs.

This was not a perfect experiment for assessing the association of monetary rewards with quest completion, as the rewards were capped at \$5000 USD, allowing a maximum of 50,000 users to claim rewards before they ran out. Further, if two quests offered monetary rewards in the same epoch, they would compete against each other until the tokens were depleted (see epoch 46 in Figure~\ref{fig:tokens}).

Despite these defects in the data, the results are insightful: even with a minimal reward, completion rates surged. The highest increase was over 12x for a quest typically completed by around 3,500 users. With the token reward, this quest (purple in Figure~\ref{fig:tokens}) saw 44,196 completions. Directly after the token reward for this quest was removed, the completion rate dropped back to its original levels before the reward was introduced. Note that in the literature on loyality programs, users prefer monetary rewards as opposed to non-monetary~\cite{RK17}.

These data reveal that the introduction of a token reward is associated with a spike in completions, while the removal of the reward corresponds with a decline back to earlier levels. These spikes suggest that token rewards do not foster loyalty or habitual engagement but rather create a purely transactional relationship. Users appear to compare the cost to the reward and only engage when there is a clear opportunity for profit.

In epoch 46, two quests with token rewards showed different outcomes. While the yellow quest (which had a low cost) saw a significant increase in completions, the red quest (with a cost higher than the reward) experienced no increase at all. If users were engaging with the platform out of loyalty or mastery, they would have completed the red quest as well as the yellow quest. However, this did not happen: users who claimed the reward for the yellow quest did not complete the red quest, leaving the reward for the red quest unclaimed.


\section{Discussion}

\subsection{DQ1: Farming and Bots } 

We currently lack data revealing how many task completions come from genuine users as opposed to farmers (humans and bots). The Fogg behavior model~\cite{Fogg09} (B=MAP) states that a behaviour requires motivation (M), ability (A), and a prompt (P). If any element is missing, the behaviour likely will not occur. Under this model, humans and bots differ: even tiny rewards can motivate (M) bots at scale (picking up every penny), while humans aware of the tasks will likely not change (either continuing completion due to loyalty or continuing to ignore them). Bots also detect prompts (P) faster than humans, including generalized MEV bots that  monitoring mempools and front-run profitable transactions~\cite{DGK+20}. This ability (A) to react instantly sets bots apart. Both farmers and bots operating from sybil addresses split tokens across many newly created addresses, each with minimal capital to cover fees, while genuine users often keep higher balances.

Putting this together, we suggest three heuristics that may help distinguish genuine users from bots: (1) If a quest alternates between unprofitable and profitable, bots will be sensitive to the exact moment it becomes marginally profitable and act; (2) if a quest requires a threshold (minimal amount of tokens), bots will be sensitive to ensuring the address meets but does not exceed the threshold amount; and (3) if a (profitable) quest is announced (especially unexpectedly), bots will be the quickest to react. We emphasize that if these heuristics were deployed in a way that harmed bots (\eg banning rather than passive measurement), bots can adjust their behaviour to evade detection, resulting in a `cat and mouse' game between farmers and quest creators. That said, we suggest new quest designs that might help identify and quantify farming activity. 

The first design observes that if costs (\ie gas fees, service fees, buying a token or NFT as part of the quest, \etc) are slightly higher than rewards (\ie tokens of monetary value received for quest completion), genuine users might be happy they are at least receiving a subsidy for a quest they would complete anyways (\cf~\cite{LdM07}), while bots will avoid unprofitable tasks. When benefits are slightly higher than costs, bots will engage (including general-purpose Maximal Extractable Value (MEV) bots~\cite{DGK+20} that agnostic about the quest system). For many quest designs, profitability could vary due to external factors, such as fluctuations in gas or token values, creating the conditions for a `natural experiment.' New engagement when quests become slightly profitable could be largely attributed to bots.

The second idea corresponds to the use of thresholds. If a quest can only be completed by addresses holding at least $X$ tokens for a certain period of time (days or weeks), bot farms need to divvy their tokens between their addresses. The transaction trace of this unusual activity could augment other patterns used by exiting farming-detection algorithms~\cite{FTWC23,ZCH+24}.


\begin{table}[t]
    \centering

    \begin{tabular}{c | c | m{3.5cm} | m{3.5cm}}
    \hline
    \textbf{Day} & \textbf{Refill Time} & \textbf{Immediate Claims (within 2 minutes)} & \textbf{Duration to Deplete Rewards } \\
    \hline
    1  & 13:00:00 & --   & 12h    \\
    2  & 12:02:17 & 200  & 1h50m  \\
    3  & 13:25:45 & 500  & 1h33m  \\
    4  & 12:09:01 & 1500 & 58m    \\
    5  & 12:45:58 & 350  & 1h30m  \\
    6  & 12:27:53 & 200  & 1h40m  \\
    7  & 13:30:17 & 1500 & 28m    \\
    8  & 16:30:07 & 300  & 2h41m  \\
    9  & 11:59:57 & 1500 & 27m    \\
    10 & 11:59:57 & $>1500$ & $<30$m \\
    11 & 11:59:57 & $>1500$ & $<30$m \\
    \hline
    \end{tabular}
        \caption{Daily Refill Times and Depletion Data.\label{tab:claims}}
    \end{table}

A final idea builds on the speed at which bots react. In the quest system we studied, a smart contract was randomly replenished with a fixed daily reward pool of \$500. Pre-enrolled participants with sufficient loyalty points could claim a nominal reward (\$0.01 or \$0.025, depending on tier) until the pool was depleted. Looking at Table~\ref{tab:claims}, we see that depletion happened quicker each day. We also see that refill times were not always predictable. On Day 7, for instance, despite a one-hour delay in the refill time, approximately 1,500 claims were made within 2 minutes. The following day, a more drastic delay of 3 hours saw claims drop to 300, likely due to the implementation of rate-limiting and a mandatory wait period between claims. Notably, this led to a temporary spike in the time required to deplete all rewards, which rose to 2h41m. By Day 9, however, the refill occurred 4.5 hours earlier, and bots had clearly adapted, with 1,500 claims occurring within 2 minutes and the time to deplete rewards dropping to 27 minutes. Finally, following the introduction of an automated daily refill schedule on Days 10 and 11, the system reached its operational minimum, with rewards being depleted in under 30 minutes and claims stabilizing at ~1,500 within the first 2 minutes.

Although these observations are suggestive, they are not conclusive for two reasons: (1) a backend system (using Cloudflare) attempted to block or rate-limit repeated bot requests, requiring multiple submissions for a single on-chain claim to succeed, and (2) the smart contract batched multiple claims into a single transaction, potentially delaying some submissions that had occurred earlier. A more direct on-chain approach---where each wallet claim corresponds to an individual transaction---would likely yield clearer data.


\subsection{DQ2: Stakeholder Analysis} 


\renewcommand{\arraystretch}{1.3}

\newcommand{\yay}{\textcolor{black}{$\Uparrow$}\xspace}
\newcommand{\nay}{\textcolor{red}{$\Downarrow$}\xspace}
\newcommand{\neu}{}
\newcommand{\con}{\textcolor{blue}{$\circlearrowleft$}\xspace}

\begin{table}[t!]

\adjustbox{}{

\begin{tabular}{p{5cm}m{0.5cm} m{0.5cm} m{0.5cm} m{0.5cm} m{0.5cm} m{0.5cm} m{0.5cm} m{0.5cm}}

\multicolumn{1}{l}{~} &
\headrow{Enhances activity} & 
\headrow{Enhances education} & 
\headrow{Enhances lock-in}  & 
\headrow{Resilient to farming} & 
\headrow{Revenue for project} & 
\headrow{Revenue for third parties} & 
\headrow{Revenue for user} \\ \hline 

\multicolumn{1}{c}{\textbf{Stakeholder}} & \multicolumn{7}{c}{\textbf{Deliverables}} \\

\hline
Project: Developers			& \yay	& \yay	& \yay	& \con	& \yay	& \neu	& \neu 	\\ \hline
Project: Executives			& \yay	& \yay	& \yay	& \nay	& \yay	& \yay	& \con 	\\ \hline
Project: Investors			& \yay	& \yay	& \yay	& \con	& \yay	& \neu	& \yay 	\\ \hline
Project: Token Holders		& \yay	& \neu	& \yay	& \nay	& \yay	& \neu	& \neu 	\\ \hline

Users: Transactional			& \neu	& \neu	& \neu	& \neu	& \neu	& \neu	& \yay 	\\ \hline
Users: Exploratory			& \neu	& \yay	& \neu	& \neu	& \neu	& \neu	& \neu 	\\ \hline
Users: Loyal                             & \yay	& \yay	& \yay	& \yay	& \neu	& \neu	& \neu 	\\ \hline

Third Party: DApps			& \yay	& \yay	& \yay	& \yay	& \neu	& \yay	& \neu 	\\ \hline
Third Party: Quest Platforms	& \yay	& \neu	& \nay	& \neu	& \neu	& \neu	& \yay 	\\ \hline
Third Party: Metric Platforms	& \neu	& \neu	& \neu	& \yay	& \neu	& \neu	& \neu 	\\ \hline

\end{tabular}
} 

\caption{A set of stakeholder groups and potential deliverables from a quest platform. Each cell indicates the priority of the stakeholder for the deliverable: empty means the priority is ambivalent to moderately favourable, \yay means it is high priority, \nay means the stakeholder opposes the deliverable, and \con indicates conflicting priorities within the stakeholder group. \label{tab:stake}}
\end{table}

\renewcommand{\arraystretch}{1.0}

A deeper question is what stakeholders actually prioritize. Is it the elimination of bots, metric boosts, revenue, other things? We conduct a \textit{stakeholder analysis}. Our stakeholder analysis (Table~\ref{tab:stake}) is based on our assessment as researchers of the preferences and incentives of different groups, rather than empirical data collected through surveys or interviews. This approach follows the methodology used by Clark et al.~\cite{CvORSZ21} for secure email systems. For space constraints, the full explanation of Table~\ref{tab:stake} is in the full version of the paper.\footnote{Appendix, Full version: \url{https://arxiv.org/abs/2501.18810}}

We highlight the column resilient to farming as the most contentious. DApps favor platforms with large genuine user bases, supporting anti-farming measures. Project-based stakeholders often tolerate farming or turn a blind eye. Investors accept inflated metrics when selling but seek real user data when buying, creating internal conflict. Developer teams face the same tension—wanting ground truth when selecting projects but valuing vanity metrics once committed.


\section{Concluding Remarks}

Many teams (based on personal conservations and social media) sense that their platform has attracted a significant number of bots. In some instances, these teams recognize that their applications or quest systems have effectively become `farming machines' for bots. Farming is a kind of ``Midas touch' for projects, immediately boosting metrics and engagement. But ultimately, it forces teams into an uncomfortable dilemma: either admit that much of their previous participation came from bots---risking tension with partners, genuine users, and investors---or remain silent and appear to lose the bulk of their user base.


\subsubsection*{Acknowledgements.} 

The authors thank the reviewers, shepherd, and workshop chairs for pointing out many issues that greatly improved the paper and its results. J. Clark acknowledges support for this research project from NSERC, Raymond Chabot Grant Thornton, and Autorité des Marchés Financiers. 


\bibliography{bib/quests.bib}
\nocite{*}


\clearpage
\appendix

\section{Further Stakeholder Analysis} 

Table~\ref{tab:stake} provides, as rows, a set of stakeholder groups and, as columns, a set of possible deliverables a quest system might provide. Deliverables are phrased to be generally favourable, so most stakeholder groups will default to being moderately in favour or, at worst, ambivalent (empty cell). Cells showcase when there is strong support, strong opposition, or internal conflict over a priority. This table helps us locate tensions between stakeholder groups. 
The first column describes the main function of a quest system: to enhance genuine user activity, a goal no stakeholder group opposes. Quest completions are not `genuine' in the sense of economic activity---quests are completed for the sake of the quest, which is superficial. However we use `genuine' to distinguish from attempts to game the quest system (which we will turn to in column 4). In the second column, again no stakeholder group opposes quests with educational value for users, while such quests are strongly preferred by exploratory users.

As of writing, the blockchain space is fragmented with many competing L1s, L2s, and services for common on-chain tasks (\eg DEXes, lending protocols, leveraged trading, \etc). Market share and user retention is a basic business goal of projects and their loyal users. Projects offering quests will typically only include tasks that engage with their own platform, while quest platforms (the `loyalty layer') offer a single hub for quests across multiple projects. Their niche is based on quests being spread across competitors. Occasionally projects will join quest platforms and then poach users for their own internal quest systems---called a `vampire attack.'

The most contentious deliverable is the elimination of farming---bots, sybils, and other methods used by transactional users to automate task completion. The basic tension is that inflated completion metrics (\eg monthly active users (MAU)) paint the picture of a fast-growing, robust user base, but its illusionary nature (`vanity metrics') could compromise credibility or collapse if transactional users exhaust the opportunities and move on to newer projects. Anti-farming measures are supported by DApps choosing a project to deploy on, favouring platforms with large genuine user bases. Joining them are metric platforms, which provide rankings of most popular projects, will lose credibility and relevance if it is easy for projects to use quests to game their metrics and climb the leaderboard. Loyal users receive more recognition when their contributions are not drowned out by a sea of fake engagement.

We expect project-based stakeholders to tacitly support farming (or at least turn a blind eye). Investors understand this when they are in a position of selling or exiting, however they want a real picture of the user base when buying into a project---thus are in internal conflict. Developer teams are similar, wanting ground truth when choosing between projects to work on and wanting success once in a team. 

When it comes to the final three columns, which emphasize the main groups that might make revenue from quests, there are no big surprises. No one opposes stakeholder capturing some revenue and stakeholder groups are self-interested, prioritizing revenue to themselves. In our evaluation, we assume revenues are non-zero sum---that is, revenue to one stakeholder group does not siphon profit from the others.

\end{document}